\documentclass[sigconf]{acmart}
\AtBeginDocument{%
  }

\copyrightyear{2025}
\acmYear{2025}

\usepackage{balance}
\usepackage{subcaption}
\usepackage{multirow}
\begin{document}

\title[ParaStyleTTS: Efficient and Robust Speaking Style Control]{ParaStyleTTS: Toward Efficient and Robust Paralinguistic Style Control for Expressive Text-to-Speech Generation}


\author{Haowei Lou}
\email{haowei.lou@unsw.edu.au}
\orcid{0009-0009-1359-872X}
\affiliation{%
  \institution{University of New South Wales}
  \city{Sydney}
  \state{NSW}
  \country{Australia}
}

\author{Hye-Young Paik}
\email{h.paik@unsw.edu.au}
\orcid{0000-0003-4425-7388}
\affiliation{%
  \institution{University of New South Wales}
  \city{Sydney}
  \state{NSW}
  \country{Australia}
}

\author{Wen Hu}
\email{wen.hu@unsw.edu.au}
\orcid{0000-0002-4076-1811}
\affiliation{%
  \institution{University of New South Wales}
  \city{Sydney}
  \state{NSW}
  \country{Australia}
}

\author{Lina Yao}
\email{lina.yao@data61.com.au}
\orcid{0000-0002-4149-839X}
\affiliation{%
  \institution{CSIRO's Data61}
  \institution{University of New South Wales}
  \city{Sydney}
  \state{NSW}
  \country{Australia}
}




\begin{abstract}
Controlling speaking style in text-to-speech (TTS) systems has become a growing focus in both academia and industry. While many existing approaches rely on reference audio to guide style generation, such methods are often impractical due to privacy concerns and limited accessibility. More recently, large language models (LLMs) have been used to control speaking style through natural language prompts; however, their high computational cost, lack of interpretability, and sensitivity to prompt phrasing limit their applicability in real-time and resource-constrained environments.
In this work, we propose ParaStyleTTS, a lightweight and interpretable TTS framework that enables expressive style control from text prompts alone. ParaStyleTTS features a novel two-level style adaptation architecture that separates prosodic and paralinguistic speech style modeling. It allows fine-grained and robust control over factors such as emotion, gender, and age. Unlike LLM-based methods, ParaStyleTTS maintains consistent style realization across varied prompt formulations and is well-suited for real-world applications, including on-device and low-resource deployment.
Experimental results show that ParaStyleTTS generates high-quality speech with performance comparable to state-of-the-art LLM-based systems while being 30x faster, using 8x fewer parameters, and requiring 2.5x less CUDA memory. Moreover, ParaStyleTTS exhibits superior robustness and controllability over paralinguistic speaking styles, providing a practical and efficient solution for style-controllable text-to-speech generation. Demo can be found at \url{https://parastyletts.github.io/ParaStyleTTS_Demo/}. Code can be found at \url{https://github.com/haoweilou/ParaStyleTTS}.
\end{abstract}

\ccsdesc[500]{Computing methodologies~Natural language processing}
\ccsdesc[500]{Computing methodologies~Supervised learning}

\keywords{Text-to-Speech, Speech Generation, Multilingual Style Adaptation, Generative Artificial Intelligence}



\maketitle

\section{Introduction}

\begin{table*}[t]
\centering
\caption{Comparison of style-controllable TTS models}
\resizebox{\textwidth}{!}{
\begin{tabular}{lcccccc}
\toprule
\textbf{Model} & \textbf{Style Control Method} & \textbf{Control Level} & \textbf{Prompt-based} & \textbf{Multilingual} & \textbf{End-To-End} & \textbf{Paralinguistic Control} \\
\midrule
StyleSpeech               & Hard tokens                   & Phoneme           &      &    &   &   \\
LanStyleTTS-Base              & Hard tokens                   & Phoneme           &      & $\checkmark$ &  &  \\
LanStyleTTS-VITS              & Hard tokens                   & Phoneme           &      & $\checkmark$ & $\checkmark$ &  \\
VITS                   & Speaker Embedding & Sentence          &  &    & $\checkmark$   \\
Spark-TTS  & Speech Prompt          & Sentence & $\checkmark$ & $\checkmark$ &  &  \\
CosyVoice  & Text Prompt          & Sentence & $\checkmark$ & $\checkmark$ &  & $\checkmark$ \\
\textbf{ParaStyleTTS (Ours)} & Hard tokens + Text Prompt & Phoneme \& Sentence & $\checkmark$ & $\checkmark$ & $\checkmark$  & $$\checkmark$$ \\
\bottomrule
\end{tabular}
}
\label{tab:model_comparison}
\end{table*}

Text-to-Speech (TTS) generation has made significant progress in recent years. It is an essential component of human-computer interaction in applications such as virtual assistants, audiobooks, and accessibility tools. Modern TTS systems aim not only to produce intelligible and natural, human-like speech but also need to support expressive and controllable generation that can generate speech with different speaking style.

Earlier TTS models such as Tacotron2~\cite{wang2017tacotron}, FastSpeech~\cite{ren2019fastspeech,ren2020fastspeech}, Glow-TTS~\cite{kim2020glow}, and VITS~\cite{kim2021conditional} focused primarily on improving intelligibility and naturalness. In particular, VITS introduces a fully end-to-end architecture that unifies the acoustic model and vocoder into a single neural network. It enhances both audio quality and generation efficiency by removing the need for external modules.

Recent advances in stylized and controllable speech generation aim to enhance the expressiveness and flexibility of TTS models.
Some works have attempted to control prosodic style variations across different languages. For instance, StyleSpeech~\cite{lou2024stylespeech} enables control tone in Chinese by disentangling tonal prosody styles during the text tokenization stage. Similarly, LanStyleTTS~\cite{lou2025generalized} proposes a similar approach to control language-specific prosody style and enables manipulation of tone and stress patterns across multiple languages.
However, beyond prosody styles, paralinguistic styles, such as emotion, age, and gender are also critical for speech generation. These factors influence how speech is perceived and are essential for personalized applications such as voice assistants, storytelling and dialogue systems with emotion.

While StyleSpeech~\cite{lou2024stylespeech} and LanStyleTTS~\cite{lou2025generalized} are effective at controlling prosodic styles, they are not well-suited for handling paralinguistic styles. Their phoneme-level fusion of style and phoneme embeddings is tailored to prosody, which affects phoneme articulation, but lacks the flexibility to model higher-level, paralinguistic-related speaking styles such as speaker's emotion, age, and gender.

Recent advances in large language models (LLMs)~\cite{yao2024survey} demonstrate strong capabilities in natural language understanding and text generation. These strengths have motivated the use of LLMs in speech generation, particularly for controlling the paralinguistic styles of speech.
CosyVoice~\cite{du2024cosyvoice} explores the use of LLMs to enable paralinguistic control in speech. 
In CosyVoice, a descriptive style prompt (e.g., "a young woman speaking angrily") is concatenated with the text input and processed by an LLM. The LLM encodes both content and style into a unified semantic embedding, which serves as conditioning for the speech decoder. This enables the model to guide speech generation based on the implied paralinguistic styles in the prompt. While this approach allows for flexible and expressive synthesis, it also introduces several limitations.

First, the speaking style and content are implicitly entangled by the LLM in an auto-regressive manner. The black-box nature of LLMs limits interpretability, making it difficult to understand or control how style is applied in the generated speech. Second, LLM-based models are computationally expensive, requiring substantial memory and inference time, which makes them unsuitable for real-time or on-device deployment. Third, the lack of explicit control and transparency reduces the robustness of the TTS system which make the style of speech highly sensitive to the phrasing of the input prompt.

To address the limitations of high computational cost and limited interpretability in LLM-based approaches.  We propose ParaStyleTTS, a lightweight, controllable, and expressive TTS framework that enables rich style control through a novel two-level style modeling architecture. Inspired by LanStyleTTS's use of prosody style tokens at phoneme level and VITS's end-to-end design, ParaStyleTTS introduces an end-to-end framework that is capable of controlling both prosodic and paralinguistic styles at the phoneme and sentence levels. Designed for end-to-end training and inference, ParaStyleTTS achieves high-quality speech generation while offering improved interpretability and computational efficiency. Key contributions of this work are as follows:
\begin{itemize}
    \item We propose a novel two-level style-controllable TTS model that explicitly disentangles prosodic and paralinguistic styles, enabling fine-grained and interpretable control over speaking style in speech synthesis.

    \item Our system is lightweight and computationally efficient, featuring an end-to-end architecture that supports expressive speech generation and is well-suited for real-time and edge-device deployment.

    \item  Extensive experiments show that ParaStyleTTS achieves robust and consistent style control across varied prompt formulations, with improved generalizability in real-world scenarios.

\end{itemize}
Experimental results show that our proposed method can generate high-quality speech with performance comparable to state-of-the-art LLM-based speech generation models while achieving 30x faster inference,  8x smaller model size, and 2.5x lower CUDA memory usage.

\begin{figure*}[t]
  \centering

  \begin{subfigure}[b]{0.55\linewidth}
    \includegraphics[width=\linewidth]{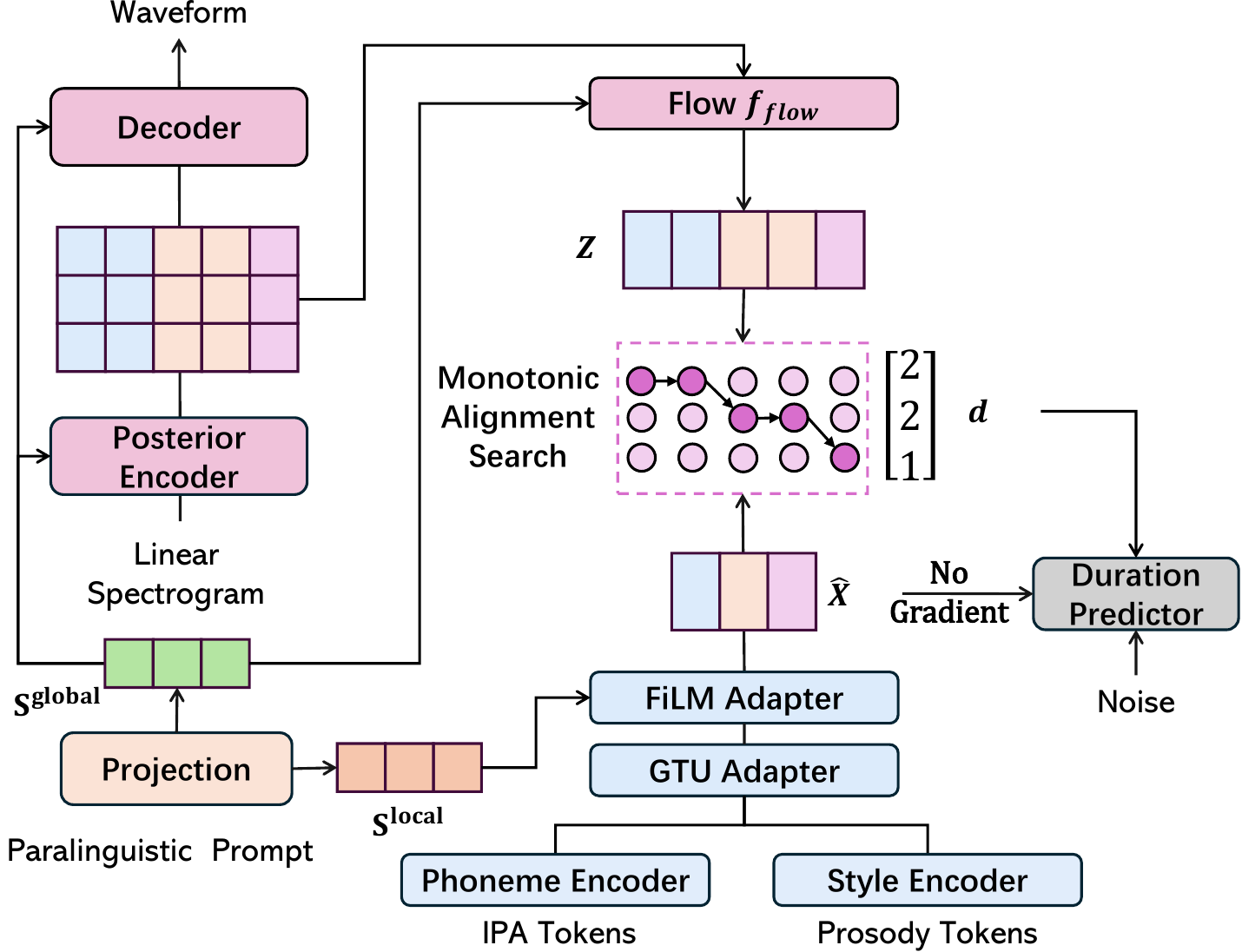}
    \caption{Training pipeline of ParaStyleTTS}
    \label{fig:ParaStyleTTS_train}
  \end{subfigure}
  \hfill
  \begin{subfigure}[b]{0.38\linewidth}
    \includegraphics[width=\linewidth]{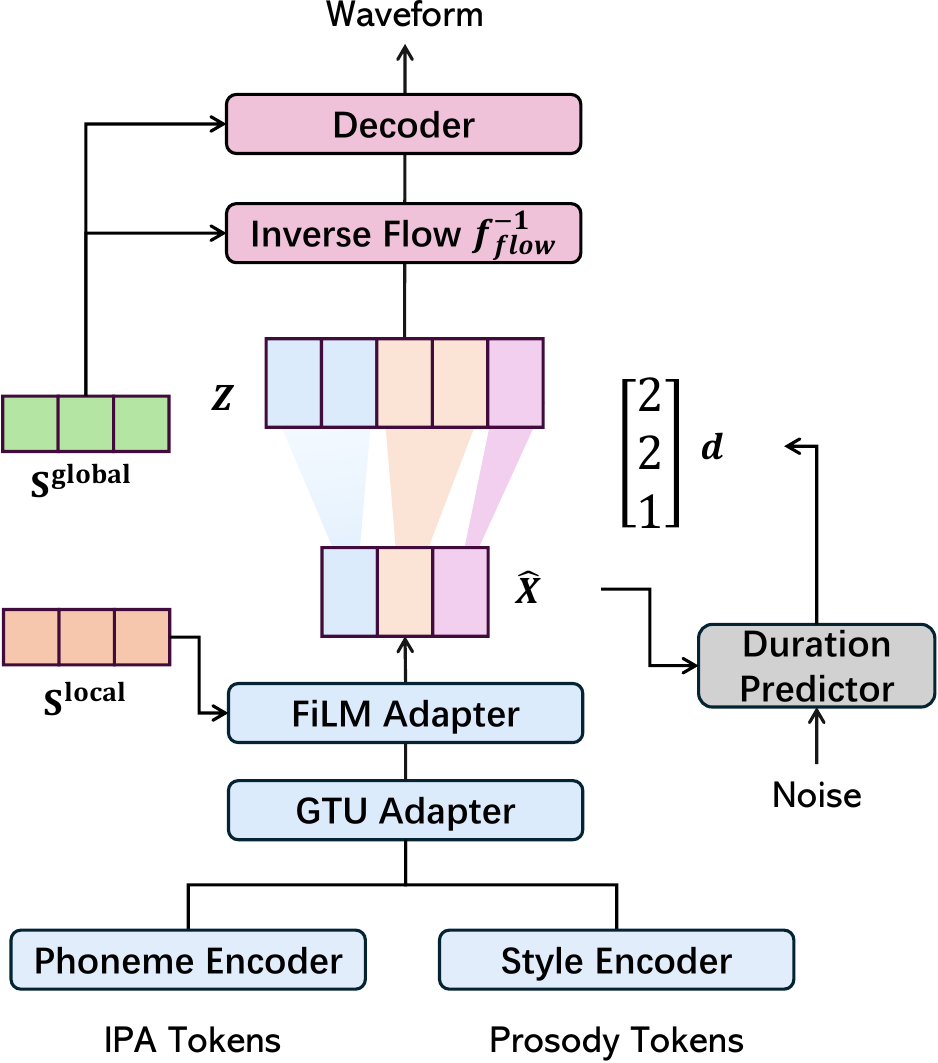}
    \caption{Inference pipeline of ParaStyleTTS}
    \label{fig:ParaStyleTTS_infer}
  \end{subfigure}
  \caption{Overall architecture of ParaStyleTTS. The system integrates text or prompt-derived paralinguistic embeddings directly into the speech generation pipeline. Unlike prior multi-stage approaches, our model encodes style prompts and phonemes jointly and generates speech in a fully end-to-end pipeline.}
  \label{fig:ParaStyleTTS_overall}
\end{figure*}

\section{Related Work}
Recent advances in style-controllable TTS models have aimed to enhance expressiveness, multilingual capabilities, and controllability over various aspects of speech such as prosody, emotion, and speaker identity. Table~\ref{tab:model_comparison} provides a comparative overview of representative models, categorized by their control method and levels of style control.

VITS~\cite{kim2021conditional} adopts a variational autoencoder framework that learns the speaking style of speakers from the training data and uses learned speaker embeddings to control the timbre and speaking style of the generated speech during inference. While it achieves high naturalness and supports direct waveform generation through an end-to-end architecture, it lacks prompt-based controllability and disentangled style modeling. Paralinguistic styles are typically entangled within the latent variables with limited interpretability and fine-grained control.

StyleSpeech~\cite{lou2024stylespeech} and LanStyleTTS~\cite{lou2025generalized} introduce phoneme-level style control mechanisms by aligning prosodic style tokens with phonemes. Each phoneme is associated with its own prosody style token, enabling fine-grained, interpretable control over prosodic features. This approach is particularly effective for tonal languages such as Chinese. However, these models are not end-to-end and rely on external vocoders, limiting their efficiency. Moreover, they struggle to generalize to

StyleSpeech~\cite{lou2024stylespeech} and LanStyleTTS~\cite{lou2025generalized} introduce phoneme-level style control mechanisms by aligning prosodic style tokens with individual phonemes. Each phoneme is associated with its own style token, enabling fine-grained and interpretable control over prosodic features. This design is particularly effective for tonal languages such as Chinese, where pitch and intonation are linguistically meaningful. However, these models lack the ability to control high-level paralinguistic styles such as emotion, age, or intent, which limits their expressiveness in broader speech generation scenarios.

More recent models, including Spark-TTS~\cite{wang2025spark} and CosyVoice~\cite{du2024cosyvoice}, leverage either speech or text-based prompt with LLMs to control speaking style. CosyVoice introduces a text-prompted system that enables paralinguistic control. Specifically, it concatenates the style prompt and content into a single input sequence and relies on LLMs to convert this sequence into meaningful semantic tokens for decoding into stylized speech. 

While this approach provides flexibility and multilingual generalization, it also introduces notable limitations. Our experiments show that CosyVoice is highly sensitive to prompt phrasing. For example, altering the prompt from “a female speaker” to “a female speaker is speaking Chinese” can lead to 
a speech generated with an incorrect speaking style. In one case, the model generated speech with a female voice even when the prompt explicitly described a male speaker. This suggests that the model may overfit to specific prompt formulations seen during training, resulting in poor generalization to compositionally complex or open-ended prompts. A more systematic evaluation is warranted to assess the robustness and reliability of prompt-based style control in such models.

\section{Method}

\subsection{Text Tokenization}\label{sec:text_tokenize}
We adopt the IPA-based text tokenization method from LanStyleTTS \cite{lou2025generalized} to convert English and Chinese text into phonemes with accompanying prosody features such as stress (English) and tone (Chinese). More specifically, each word in English is converted to ARPAbet phonemes using the CMU Pronouncing Dictionary~\cite{cmu_pronouncing_dict}, with stress markers extracted separately to form the style token sequence. The phonemes are then mapped to IPA phonemes.
For Chinese, we use the \texttt{pypinyin} library to convert each character into its Pinyin form, which is then split into initials and finals. The tone is stripped from the final and used as a style token, while the remaining components are mapped to IPA phonemes.

Our IPA dictionary comprises 81 phonemes, 5 tone markers for Chinese, and 3 stress markers for English. In addition, we include three special tokens: \texttt{[START]} to denote the beginning of a sequence, \texttt{[END]} for the end of a sequence, and \texttt{[|]} as a word boundary marker.

\subsection{Token Encoder}
We first project the IPA tokens~$X$ and prosody style tokens~$S$ into a sequence of vector representations using two distinct embedding layers. To preserve sequential information, sinusoidal positional encodings are added to the embeddings. Then, IPA and prosody style embeddings are fed into Feed-Forward Transformer (FFT) blocks~\cite{ren2019fastspeech} to leverage the self-attention to model long-range dependencies and contextual relationships across the token sequence. Unlike the original Transformer architecture~\cite{vaswani2017attention}, which uses a two-layer fully connected network in its feed-forward submodule, we replace it with two one-dimensional convolutional layers to better capture local contextual dependencies between adjacent tokens.

Let $\mathbf{X} = [x_1, x_2, \ldots, x_L]$ denote the phoneme embedding sequence obtained from text tokenization. We associate each phoneme $x_l$ with a corresponding prosodic style embedding $\mathbf{x}_t,\mathbf{s}^{\text{pho}}_t \in \mathbb{R}^{d_1}$, forming the phoneme-level prosody sequence:
\[
\mathbf{S}^{\text{pho}} = [\mathbf{s}^{\text{pho}}_1, \mathbf{s}^{\text{pho}}_2, \ldots, \mathbf{s}^{\text{pho}}_L]
\] 

\subsection{Paralinguistic Encoder}
The sentence-level paralinguistic style embedding is denoted as $\mathbf{S}^{\text{para}} \in \mathbb{R}^{d_2}$, representing sentence-level paralinguistic characteristics such as emotion, age, gender, and accent. To obtain this embedding, we employ a pre-trained MPNet model~\cite{song2020mpnet} to encode descriptive paralinguistic prompts into $d_2$-dimensional embedding.
For each speech sample in our dataset, we construct a text prompt using the following template:
\begin{quote}
\textit{"A [Age] [Gender] is speaking [Accent] with [Emotion] emotion.”}
\end{quote}
It is then fed into MPNet to produce the paralinguistic prompt embedding $\mathbf{S}^{\text{para}}$, which is used to condition the TTS model and guide the generation of speech with the intended paralinguistic style.

\subsection{Style Adapter}
Style in speech generation is a broad concept that encompasses both prosodic features, such as pitch and tone, and paralinguistic styles, including gender, emotion, accent, and more. In this research, we divide style into two categories with different levels of control. One is the phoneme-level style, which captures fine-grained prosodic variations such as tone and stress at the level of individual phonemes. This level strongly influences how each word is articulated. Another one is sentence-level style, which represents global characteristics of the speech. It includes emotion, age, gender, and accent. While these features shape the overall impression of the speech, they exert less direct influence on phoneme realization. To support effective control at both levels, we design specialized architectures tailored to the unique requirements of each level of style.

\subsubsection{Prosody Style Adapter}
Given the phoneme embedding sequence $\mathbf{X} = [x_1, x_2, \ldots, x_L]$ and the phoneme-level prosody style sequence $\mathbf{S}^{\text{pho}} = [\mathbf{s}^{\text{pho}}_1, \mathbf{s}^{\text{pho}}_2, \ldots, \mathbf{s}^{\text{pho}}_L]$, we apply a lightweight adapter to inject prosodic features into the phoneme representations. Following the design in LanStyleTTS~\cite{lou2025generalized}, we employ a Gated Tanh Unit (GTU) fusion mechanism, defined as:

\begin{equation}
\tilde{\mathbf{x}}_t = \tanh(W_1 x_t + b_1) \odot \sigma(W_2 \mathbf{s}^{\text{pho}}_t + b_2), \quad \forall t \in \{1, \ldots, L\},
\end{equation}

where $W_1$, $W_2$ are learnable projection weights, $b_1$, $b_2$ are biases, $\odot$ denotes element-wise multiplication, and $\sigma(\cdot)$ is the sigmoid function. This formulation allows the prosody style to modulate the phoneme representation at a fine-grained level while preserving phonetic structure.

\subsubsection{Paralinguistic Style Adapter}
Given that paralinguistic style typically remains consistent throughout a speech, it can influence both phoneme-level and sentence-level acoustic characteristics. To capture these effects, we first apply two distinct linear layers to project paralinguistic prompt embedding $\mathbf{S}^{\text{para}} \in \mathbb{R}^{d_2}$ into phoneme-level~($\mathbf{S}^{\text{local}}$) and sentence-level~($\mathbf{S}^{\text{global}}$) paralinguistic style embedding.

\begin{equation}
\mathbf{S}^{\text{local}} = W_{\text{local}} \mathbf{S}^{\text{para}} + b_{\text{local}}, \quad
\mathbf{S}^{\text{global}} = W_{\text{global}} \mathbf{S}^{\text{para}} + b_{\text{global}},
\end{equation}

We adopt Feature-wise Linear Modulation (FiLM)~\cite{perez2018film} to inject phoneme-level paralinguistic style embedding into the phoneme embeddings via sequence-wise conditioning. Specifically, given the projected style embedding $\mathbf{s}^{\text{sent}}$, we compute the scaling and bias vectors as:

\begin{equation}
\gamma = W_\gamma \mathbf{S}^{\text{local}} + b_\gamma, \quad \beta = W_\beta \mathbf{s}^{\text{local}} + b_\beta,
\end{equation}

where $W_\gamma, W_\beta \in \mathbb{R}^{d_1 \times d_2}$ are learnable projection matrices, and $b_\gamma, b_\beta \in \mathbb{R}^{d_1}$ are bias terms. These parameters generate the modulation factors used to transform the phoneme embeddings:

\begin{equation}
\hat{\mathbf{x}}_t = \gamma \odot \tilde{\mathbf{x}}_t + \beta, \quad \forall t \in \{1, \ldots, L\},
\end{equation}

where $\odot$ denotes element-wise multiplication. This FiLM-based adapter integrates paralinguistic style into each phoneme within the speech.


The sentence-level paralinguistic style embedding ($\mathbf{s}^{\text{global}}$) is applied in both training and inference stages to guide sentence-level style adaptation within the waveform decoder. By conditioning on this embedding, ParaStyleTTS is able to impose consistent paralinguistic styles throughout the speech.

\subsection{Latent Embedding Learning}\label{sec:latent_embed}
We adopt the variational autoencoder (VAE) framework~\cite{kingma2019introduction} with adversarial training~\cite{goodfellow2014generative} and normalizing flows~\cite{papamakarios2021normalizing} to model expressive latent representations and decode our waveform decoder due to its fully end-to-end training, non-autoregressive inference, and high-fidelity speech generation. 

In our system, the decoder takes as input the style-integrated phoneme representations $\hat{\mathbf{X}} = [\hat{x}_1, \hat{x}_2, \ldots, \hat{x}_L]$, which are modulated by both phoneme-level prosody and sentence-level paralinguistic style. To enforce global consistency, the sentence-level embedding $\mathbf{S}^{\text{global}}$ is concatenated with both the prior and posterior encodings before being passed through the normalizing flow layers.

The latent embedding $\mathbf{Z} \in \mathbb{R}^{N\times T}$ is first sampled from a Gaussian posterior using a variational autoencoder (VAE). The posterior distribution is conditioned on the ground-truth spectrogram $\mathbf{Y}$ and the sentence-level style embedding $\mathbf{S}^{\text{global}}$, and is parameterized as:

\begin{equation}
\mathbf{Z} \sim q(\mathbf{Z} | \mathbf{Y}, \mathbf{S}^{\text{global}}) = \mathcal{N}(\boldsymbol{\mu}_{\text{post}}, \boldsymbol{\sigma}_{\text{post}}),
\end{equation}

where $\boldsymbol{\mu}_{\text{post}}, \boldsymbol{\sigma}_{\text{post}}$ are predicted by a posterior encoder from the spectrogram and global style embedding.
To obtain a more expressive latent representation, we further apply a sequence of invertible normalizing flows to $\mathbf{z}$:

\begin{equation}
\mathbf{Z}_{\text{flow}} = f_{\text{flow}}(\mathbf{Z}; \theta_{\text{flow}}),
\end{equation}

The prior distribution is defined over this transformed latent space and is modeled as a Gaussian conditioned on the style-integrated phoneme sequence $\hat{\mathbf{X}}$:

\begin{equation}
p(\mathbf{Z}_{\text{flow}} | \hat{\mathbf{X}}) = \mathcal{N}(\boldsymbol{\mu}_{\text{prior}}, \boldsymbol{\sigma}_{\text{prior}}).
\end{equation}

where \( \boldsymbol{\mu}_{\text{prior}}, \boldsymbol{\sigma}_{\text{prior}} \) are predicted by a prior encoder network, which takes as input the phoneme embeddings $\tilde{\mathbf{X}}$ and \textbf{local} style embedding $\mathbf{S}^{\text{local}}$. These parameters define the expected distribution of latent speech features given the linguistic content and phoneme-level paralinguistic style. The KL-divergence~\cite{kingma2013auto} between the transformed posterior sample and the prior is minimized during training:

\begin{equation}
\mathcal{L}_{\text{KL}} = D_{\text{KL}}\left(\mathbf{Z_{\text{flow}}} \parallel p(\mathbf{Z}_{\text{flow}} | \hat{\mathbf{X}})\right).
\end{equation}

This structure allows the model to capture a rich and flexible latent distribution that aligns with both the local and global style information.

\subsection{Duration Alignment and modeling}
To align the phoneme sequence integrated with paralinguistic style embedding $\hat{\mathbf{X}} = [\hat{x}_1, \hat{x}_2, \ldots, \hat{x}_L]$ with the latent embedding~$\mathbf{Z}$ during training, we apply Monotonic Alignment Search (MAS)~\cite{kim2021conditional} to compute a soft alignment matrix \( \mathbf{A} \in \mathbb{R}^{L \times T} \), where \( A_{t, j} \) represents the attention weight between the $t$-th phoneme and the $j$-th frame in $\mathbf{Z}$. 

The duration \( d_t \) for each phoneme is estimated by summing the attention weights over the time axis:

\begin{equation}
d_t = \sum_{j=1}^{T} A_{t, j}, \quad \forall t \in \{1, \ldots, L\}.
\end{equation}

We integrate a Stochastic Duration Predictor (SDP) to learn to predict the log-duration distribution conditioned on the phoneme and style features. During training, we minimize a log-domain Mean Squared Error (MSE) loss between the predicted and reference durations:

\begin{equation}
\mathcal{L}_{\text{dur}} = \frac{1}{L} \sum_{t=1}^{L} \left( \log(d_t + \epsilon) - \log(\hat{d}_t + \epsilon) \right)^2,
\end{equation}
where \( \epsilon \) is a constant for numerical stability.

It allow the model to intrinsically learn phoneme durations during training, while also capturing duration variations influenced by paralinguistic styles. 

\subsection{Training Objective}
The model is optimized using a combination of objectives adapted from the VITS framework~\cite{kim2021conditional}. These include a reconstruction loss $\mathcal{L}_{\text{recon}}$, which measures the difference between the generated~$\hat{Y}$ and ground-truth spectrogram, an adversarial loss~$\mathcal{L}_{\text{adv}}$ to encourage realistic waveform generation through multi-period discriminators $D$ and a feature matching loss~$\mathcal{L}_{\text{fm}}$ to stabilize adversarial training by aligning discriminator's internal feature of real and generated speech:

\begin{align}
\mathcal{L}_{\text{recon}} &= \| \mathbf{Y} - \hat{\mathbf{Y}} \|_1  \\
\mathcal{L}_{\text{adv}} &= \mathbb{E}_{\hat{\mathbf{Y}}}[(D(\hat{\mathbf{Y}}) - 1)^2] \\
\mathcal{L}_{\text{fm}} &= \sum_{l=1}^{L} \| D^{(l)}(\mathbf{Y}) - D^{(l)}(\hat{\mathbf{Y}}) \|_1 \\
\mathcal{L_{\text{total}}} &= \mathcal{L_\text{fm}} + \mathcal{L_\text{KL}}  + \mathcal{L_\text{dur}}  + \mathcal{L_\text{recon}}  + \mathcal{L_\text{adv}}
\end{align}

\subsection{Time Complexity Analysis}
To analyze the computational complexity of the overall architecture in ParaStyleTTS versus LLM-based paralinguistic style control models, we define $N$ as the length of the text sequence and $M$ as the length of the paralinguistic style prompt.

In ParaStyleTTS, the phoneme and prosody style tokens are encoded separately using transformer-based FFT blocks, followed by a Gated Tanh Unit (GTU) style adapter. The time complexity is $\mathbf{O}(N^2)$. Meanwhile, the paralinguistic style prompt is independently processed by a transformer-based MPNet encoder, contributing an additional $\mathbf{O}(M^2)$ in computation. In total, this results in a combined time complexity of $\mathbf{O}(N^2 + M^2)$.
In contrast, LLM-style fusion approaches concatenate the text and paralinguistic tokens into a single sequence of length $N + M$, which is jointly encoded by a large transformer or LLM model. This yields a total time complexity of $\mathbf{O}((N + M)^2)$.
As a result, LLM-style fusion introduces an additional cross-attention cost of $\mathbf{O}(NM)$, making it less computationally efficient.
\label{tab:overall_performance_comparision}
\section{Experiment}
\subsection{Datasets}
We conduct our experiments using a multilingual and multi-style speech corpus comprising both English and Chinese speech samples. To ensure broad coverage of paralinguistic styles, we construct a composite dataset by combining several publicly available speech datasets.
The final training data consists of four sources. The \textbf{Baker} dataset~\cite{BakerDataset2020} is a single-speaker Mandarin Chinese corpus featuring a female voice. The \textbf{LJSpeech} dataset~\cite{ljspeech17} is a single-speaker English corpus, also featuring a female speaker, widely used in TTS research for clean and consistent English utterances. The \textbf{Emotional Speech Dataset (ESD)}~\cite{zhou2021seen} contributes a multilingual, multi-speaker emotional speech in both English and Chinese, covering five emotional categories and including both male and female speakers. To further enrich stylistic diversity, we curate 16 stylized character speech captions from the \textbf{Genshin Impact} voice dataset to capture expressive speech across different age styles.

It covers \textbf{two languages} (English and Chinese), \textbf{two genders} (male and female), and \textbf{four age categories} (child, teenager, young adult, and adult). Emotion labels span \textbf{five classes}: neutral, happy, sad, angry, and surprised. 
The final training dataset has 86k speech samples, with around 108.30 hours of training data. The dataset contains speech from 38 speakers. 
More details about our dataset can be found in table~\ref{tab:dataset} in the appendix. 

\subsection{Preprocessing}
All speech recordings are resampled to 22.05 kH. To ensure the phoneme remains consistent across multilingual languages, we apply IPA-based phoneme tokenization uniformly across both English and Chinese using the text tokenization method described in Section~\ref{sec:text_tokenize}. For each speech-text pair, we pre-compute phoneme tokens and prosody-style tokens to serve as input to the TTS model. In addition, we generate a paralinguistic style caption (e.g., "A young female is speaking English with happy emotion”), which is then encoded into a style embedding to guide paralinguistic style control. 

\subsection{Training}
The model is trained on four NVIDIA V100 GPUs with a batch size of 32 for up to 700k steps. We adopt the AdamW optimizer\cite{loshchilov2017decoupled}, using the same hyperparameters and learning rate schedule as VITS\cite{kim2021conditional}. 
\begin{table}[t]
\centering
\caption{Overall Performance Comparison}\label{tab:comparison}
\begin{tabular}{l|ccc}
\toprule
Model & \textbf{WER} & \textbf{I-MOS} & \textbf{N-MOS}
\\
\midrule
StyleSpeech               &  42.33 $\pm$ 28.80 & 2.18 $\pm$ 0.30 & 1.94 $\pm$ 0.37 \\
LanStyleTTS-Base          & 21.20 $\pm$ 16.45  & 2.59 $\pm$ 0.30 & 2.24 $\pm$ 0.27  \\
LanStyleTTS-VITS          & 12.74 $\pm$ 13.11 & 4.57 $\pm$ 0.48 & 4.23 $\pm$ 0.50 \\
VITS                      & 19.32 $\pm$ 15.88 & 4.28 $\pm$ 0.38 & 3.82 $\pm$ 0.78 \\
Spark-TTS                 & 15.12 $\pm$ 14.43 & 4.45 $\pm$ 0.40 & 4.33 $\pm$ 0.49  \\
CosyVoice                 & 10.30 $\pm$ 13.52 & 4.75 $\pm$ 0.22 & 4.57 $\pm$ 0.29  \\
\hline
\textbf{ParaStyleTTS} & 15.29 $\pm$ 14.41 & 4.65 $\pm$ 0.32 & 4.36 $\pm$ 0.49 \\
\bottomrule
\end{tabular}

\end{table}
\subsection{Experiment Setting}
To evaluate the intelligibility and paralinguistic expressiveness of our system, we adopt both objective metrics and human perceptual testing.
For intelligibility assessment, we avoid using samples from our training set to ensure fair cross-model comparisons, as different baselines are trained on distinct datasets. Instead, we generate speech for two standardized corpora: the 720 Harvard Sentences\cite{institute1969ieee} in English and 400 phonetically balanced Mandarin TMNews sentences set from~\cite{chen2023baspro}, both widely recognized for evaluating speech systems. The generated speech is transcribed using Whisper-Base~\cite{radford2023robust}, and Word Error Rate (WER) is computed by comparing the transcriptions against the ground-truth text. Lower WER scores indicate higher intelligibility and transcription consistency.

To complement the objective evaluation, we conduct a Mean Opinion Score (MOS) listening test. For each language, we randomly select ten generated samples from each model and present them to at least five bilingual listeners fluent in both English and Chinese. The listeners are asked to rate each sample along two dimensions: intelligibility (I-MOS), which reflects how easily the speech content can be understood, and naturalness (N-MOS), which reflects how human-like and fluent the speech sounds.

To assess the expressiveness of paralinguistic styles, we use open-source speech analysis models to determine whether the generated speech is distinguishable by classifiers. For emotion evaluation, we adopt Emotion2Vec~\cite{ma2023emotion2vec}, a state-of-the-art, open-source model for emotion recognition in speech. We conduct classification experiments on generated samples with varying emotional labels to verify their perceptual separability.
For age and gender analysis, due to the lack of large-scale, open-source models, we train a lightweight paralinguistic style classifier based on CLAP~\cite{elizalde2023clap}. This model evaluates whether age and gender styles in the generated speech can be reliably identified.

In addition to evaluating speech quality and expressiveness, we assess inference efficiency, model size, and CUDA memory usage across different TTS models to determine their suitability for edge device deployment. All performance measurements are conducted on a single NVIDIA 3060 Ti GPU, a widely available consumer-grade device selected to reflect realistic and cost-effective deployment scenarios. To ensure reliability, we generate the entire evaluation set comprising 1,120 sentences, including the 720 Harvard Sentences and 400 phonetically balanced Mandarin sentences. Each sentence is processed individually with a batch size of one. The average inference time and peak CUDA memory usage per sample are recorded. These metrics provide a consistent and fair basis for comparing computational efficiency across models.

\section{Result \& Discussion}\label{sec:result}
In this section, we evaluate the performance of ParaStyleTTS in terms of speech quality, resource usage, speaking style controllability, and robustness.
Our research is guided by the following three research questions (RQs):
\begin{itemize}
\item \textbf{RQ1:} Can ParaStyleTTS achieve more expressive control over speaking styles?
\item \textbf{RQ2:} Can ParaStyleTTS provide effective style control in a lightweight and resource-efficient manner?
\item \textbf{RQ3:} Can ParaStyleTTS maintain robust style control under varying prompt formulations?
\end{itemize}

\subsection{Overall Comparison}
Table~\ref{tab:comparison} compares the speech quality generated by ParaStyleTTS with that of other models.
ParaStyleTTS closely matches the perceptual quality of CosyVoice and outperforms Spark-TTS and other non-LLM-based baselines. It achieves an intelligibility MOS of 4.64 and a naturalness MOS of 4.36, ranking as the second-best model in subjective evaluations.

\subsection{Speaking Style Controllability}
\begin{table}[t]
\centering
\caption{Speaking Style Expressiveness Comparison}
\begin{tabular}{l|ccc}
\toprule 
\textbf{Model} & \textbf{Emotion Acc} & \textbf{Gender Acc} & \textbf{Age Acc}
\\
\midrule
CosyVoice                 & 47.50\% & 75.00\% & 21.88\%   \\

\textbf{ParaStyleTTS (Ours)} & \textbf{54.00\%} & \textbf{100.00\%} & \textbf{57.50\%}  \\
\bottomrule
\end{tabular}

\label{tab:para_performance_comparision}
\end{table}
Table~\ref{tab:para_performance_comparision} presents a comparison of expressive speaking style control across different models. The results show that ParaStyleTTS surpasses CosyVoice in all evaluated paralinguistic dimensions, including emotion, gender, and age control. Specifically, speech generated by ParaStyleTTS achieves classification accuracies of 54.00\% for emotion, 100.00\% for gender, and 57.50\% for age. In contrast, CosyVoice obtains only 47.50\%, 75.00\%, and 21.88\% for the same categories, respectively. 

\begin{figure*}[ht]
    \centering
    \begin{subfigure}[b]{0.27\textwidth}
        \centering
        \includegraphics[width=\linewidth]{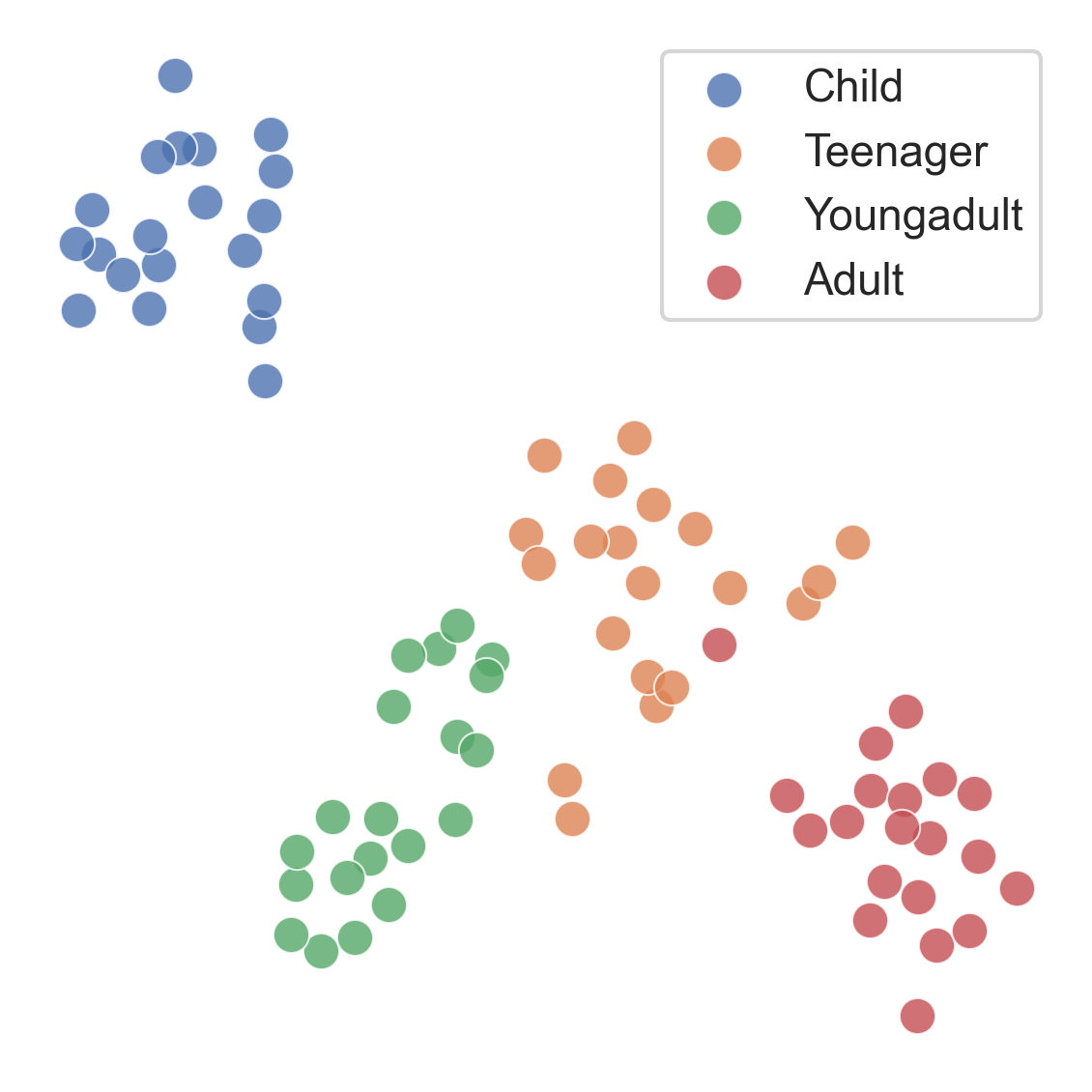}
        \caption{Age}
        \label{fig:embed-age}
    \end{subfigure}
    \hfill
    \begin{subfigure}[b]{0.27\textwidth}
        \centering
        \includegraphics[width=\linewidth]{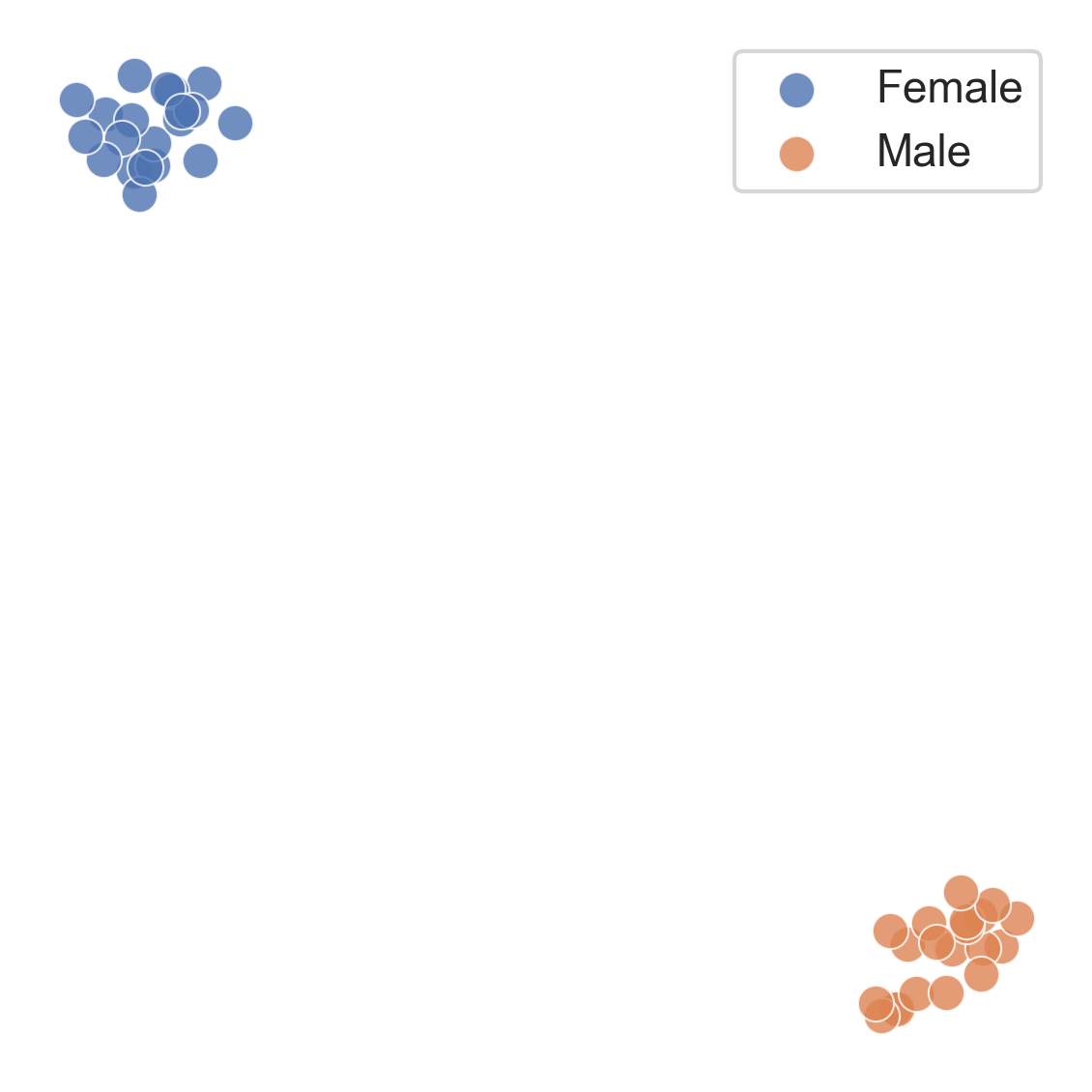}
        \caption{Gender}
        \label{fig:embed-gender}
    \end{subfigure}
    \hfill
    \begin{subfigure}[b]{0.27\textwidth}
        \centering
        \includegraphics[width=\linewidth]{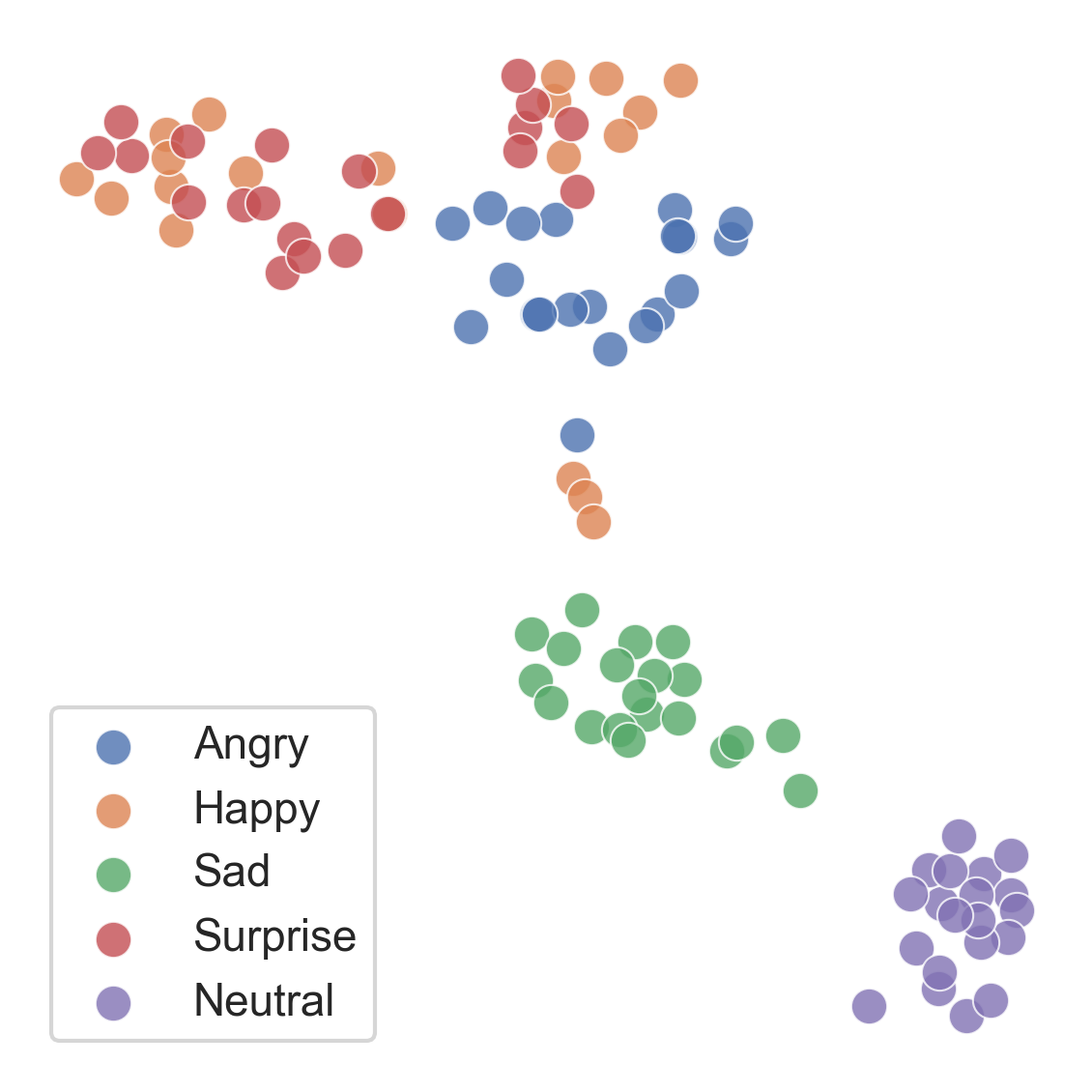}
        \caption{Emotion}
        \label{fig:embed-emotion}
    \end{subfigure}
    
    \caption{t-SNE visualizations of the speaking style embeddings extracted from speech generated by different prompts. Each dot corresponds to a speech sample.}
    \label{fig:embed-all}
\end{figure*}
To further evaluate the distinguishability of generated styles, we train a speaking style classifier and extract embeddings from ParaStyleTTS-generated speech conditioned on different style prompts.  These embeddings are visualized using t-SNE in Figure~\ref{fig:embed-all}, providing an external perspective on how well the model encodes paralinguistic styles in the acoustic space.

In Figure~\ref{fig:embed-age}, age-related embeddings form well-separated clusters, with only minor overlap between Teenagers and Young Adults. This is expected, as teenagers and young adults are relatively close in age and therefore tend to share similar vocal characteristics. Figure~\ref{fig:embed-gender} shows clearly distinct clusters for Male and Female, which confirms that the model captures gender-specific acoustic features. In Figure~\ref{fig:embed-emotion}, most emotion classes form coherent and distinguishable clusters. However, Happy and Surprise overlap noticeably, likely because both involve elevated pitch, faster speech, and high energy. These similarities reduce the model's ability to distinguish them.
By contrast, Sad, Neutral, and Angry are easier to separate. Each of these emotions shows more distinct acoustic patterns, such as slower pace, flat intonation, or sharper articulation.

These results demonstrate that ParaStyleTTS can successfully control the speaking style of generated speech. It outperforms the CosyVoice across all three evaluated paralinguistic styles and produces embeddings with high distinguishability.

This improvement can be attributed to ParaStyleTTS’s dedicated two-level style modeling architecture and its end-to-end training design.
Unlike CosyVoice, which relies on large language models (LLMs) to infer and apply speaking styles from a semantic perspective, ParaStyleTTS adopts an acoustic-centric learning paradigm. In CosyVoice, style control is driven by semantics understanding.  LLMs interpret the meaning of style prompts like happy or angry based on semantics meaning and then rely on a vocoder to generate speech. The overall process follows a pipeline from text to semantics, and from semantics to acoustics, where the acoustic characteristics of different speaking styles are modeled only indirectly.

In contrast, ParaStyleTTS learns speaking styles directly from acoustic features through supervised training with style prompts. As shown in Section~\ref{sec:latent_embed}, the latent embeddings are explicitly conditioned on style prompts, enabling the model to form direct associations between each prompt and its corresponding acoustic characteristics. This approach allows for more accurate and fine-grained control over style expression.
The two-level architecture further strengthens this capability by applying the style prompt at both the phoneme level (capturing prosody and speech rate) and the sentence level (capturing broader attributes such as emotion and age). This design enables ParaStyleTTS to generate speech that is 
 acoustically consistent with the intended speaking style.

\begin{table*}[ht]
\centering
\caption{Computational Resource Comparison}
\begin{tabular}{l|c|c|c}
\toprule
\textbf{Model} & Inference Time (ms) & Parameter Size (M) & CUDA Memory Usage (MB)\\
\midrule
StyleSpeech               & 55.8 & 63.61 & 463  \\
LanStyleTTS-Base          & 53.5 & 61.34 & 480   \\
LanStyleTTS-VITS          & 100.0 & 42.52 & 305   \\
VITS                      & 99.9 & 36.30 & 340   \\
Spark-TTS                 & 7,999.0 & 506.63 & 3,854 \\
CosyVoice                 & 4,076.1 & 436.05 & 1,852  \\
\hline
\textbf{Prompt Encoder} & 20.6 & 109.49 & 636 \\
\textbf{ParaStyleTTS (Ours)} & 121.2 & 52.51  & 763\\
\bottomrule
\end{tabular}
\label{tab:resource_comparision}
\end{table*}

\subsection{Resource usage}
Table~\ref{tab:resource_comparision} presents a resource-based comparison of TTS models in terms of inference speed, model size, and CUDA memory usage, which are three key metrics that affect real-time responsiveness, storage cost, and hardware requirements.  All models are evaluated on a single NVIDIA RTX 3060 Ti GPU. The reported results correspond to generating a speech segment approximately two seconds in duration.

As shown in the table, ParaStyleTTS demonstrates significant resource efficiency. When the prompt encoder is included, the model requires 140 ms of inference time, has 150 million parameters, and uses 763 MB of CUDA memory. However, due to the design of ParaStyleTTS, the prompt encoder can be decoupled during inference by precomputing and caching the style embeddings in a production environment. Without the prompt encoder, the runtime model size and inference time are reduced to just 52 million parameters and 121 ms, respectively. 

CUDA memory usage remains unchanged at 763 MB because the reported value reflects peak memory usage at any given point during inference. In our setup, the prompt encoder and the main ParaStyleTTS model run sequentially, not in parallel. Since they do not overlap in execution, the total memory usage at any one time is determined by the larger of the two. Because ParaStyleTTS uses more memory than the prompt encoder, the overall peak memory remains at 763 MB.

In contrast, LLM-based TTS models such as CosyVoice and Spark-TTS are significantly more resource-intensive. Among them, CosyVoice is the most lightweight, yet it still requires over 4000 ms of inference time, more than 400 million parameters, and 1852 MB of CUDA memory. Compared to CosyVoice, ParaStyleTTS achieves over \textbf{30x} faster inference, up to \textbf{8x} smaller model size, and \textbf{2.5x} lower CUDA memory usage.

These results highlight ParaStyleTTS's advantage in terms of computational efficiency. Its lightweight design enables the generation of high-quality speech while remaining highly suitable for real-time and resource-constrained environments.

This efficiency comes from the end-to-end and efficient design of ParaStyleTTS.
LLM-based approaches, such as CosyVoice, rely on a multi-stage pipeline for speech generation. Beyond using LLMs for style fusion, CosyVoice also requires a flow-matching-based vocoder~\cite{lipman2022flow} to convert the LLM output into a waveform. Each of these components adds to the overall computational load and system complexity.
ParaStyleTTS, on the other hand, leverages a fully end-to-end architecture that directly generates waveforms with no need for any additional modules. Hence, it reduces both latency and space consumption.

Another major advantage of ParaStyleTTS is that it does not rely on LLMs for style adaptation. LLMs are computationally expensive due to their large parameter sizes and nature of autoregressive processing. ParaStyleTTS avoids this overhead by directly learning the mapping between style prompts and corresponding acoustic characteristics. With the help of a lightweight style adapter, it achieves high flexibility and fast generation while maintaining expressive control over speaking style.

In addition to empirical improvements in runtime performance, ParaStyleTTS is also more efficient in terms of computational complexity. It processes phoneme tokens and style prompts independently using transformer encoders. The total time complexity is $O(N^2 + M^2)$, where $N$ and $M$ represent the lengths of the phoneme sequence and style prompt, respectively. 
This complexity can be further reduced during inference by precomputing and caching the style prompt embedding. Since the prompt no longer needs to be processed at runtime, the term associated with $M$ is eliminated. The overall complexity can be further reduced to $O(N^2)$.
In contrast, LLM-based models concatenate the phoneme and prompt tokens together and process them jointly. It results in a higher time complexity of $O((N + M)^2)$. This complexity gap becomes increasingly significant with longer inputs or richer style prompts. 
By combining architectural simplicity with both theoretical and practical efficiency, ParaStyleTTS offers a scalable and deployable solution suitable for real-time, cloud-based, and on-device TTS applications.

\subsection{Robustness Style Control}
Robust style control is critical for real-world TTS applications, where prompts may vary in phrasing but still intend to express the same speaking style. A robust model must consistently generate speech that matches the intended paralinguistic style, regardless of how the prompt is formulated.

\begin{table}[t]
\centering
\caption{Emotion Per-Class Accuracy Comparison}
\begin{tabular}{l|ccccc}
\toprule
\textbf{Model} & \textbf{Happy} & \textbf{Angry} & \textbf{Sad} & \textbf{Neutral} & \textbf{Surprise} \\
\midrule
CosyVoice          & 62.50 & 30.00 & \textbf{57.50} & \textbf{82.50} & 5.00\%    \\
\textbf{ParaStyleTTS} & \textbf{72.50} & \textbf{55.50} & 45.00 & 70.00 & \textbf{27.50 }\\
\bottomrule
\end{tabular}
\label{tab:emotion_acc}
\end{table}

\begin{table}[t]
\centering
\caption{Age Per-Class Accuracy Comparison}
\begin{tabular}{l|ccccc}
\toprule
\textbf{Model} & \textbf{Child} & \textbf{Teenager} & \textbf{YoungAdult} & \textbf{Adult} \\
\midrule
CosyVoice          & 0.00   & 42.50    & 35.00    & 10.00      \\
\textbf{ParaStyleTTS} & \textbf{82.50} & \textbf{72.50} & \textbf{50.00} & \textbf{25.00} \\
\bottomrule
\end{tabular}
\label{tab:age_acc}
\end{table}

\begin{table}[b]
\centering
\caption{Gender Per-Class Accuracy Comparison}
\begin{tabular}{l|ccccc}
\toprule
\textbf{Model} & \textbf{Male} & \textbf{Female}  \\
\midrule
CosyVoice          & 50.00   & 100.00       \\
\textbf{ParaStyleTTS} & \textbf{100.00} & \textbf{100.00} \\
\bottomrule
\end{tabular}
\label{tab:gender_acc}
\end{table}

We observe a clear gap in style accuracy and robustness between models in Tables~\ref{tab:emotion_acc}, \ref{tab:age_acc}, and \ref{tab:gender_acc}. CosyVoice frequently fails to produce speech aligned with the intended emotion, age, or gender, especially for underrepresented or subtle styles. For instance, CosyVoice achieves only 5.00\% accuracy for Surprise, 0.00\% for Child, and 50.00\% for Male. These inconsistencies indicate that its style control is fragile and often fails to match the prompted speaking style.

To further investigate robustness against prompt variation, we conduct a controlled experiment using gender as a proxy attribute. Gender is ideal for this task due to its perceptual clarity and ease of evaluation. We design a set of prompts with varied phrasing but identical semantic meaning (e.g., "A male speaker is talking", "A man is talking", etc.) and use them to guide speech generation for both ParaStyleTTS and CosyVoice. Table~\ref{tab:text_prompt} in the appendix section shows the full details of evaluated prompts.

\begin{figure}[ht]
    \centering
    \begin{subfigure}[b]{0.22\textwidth}
        \centering
        \includegraphics[width=\linewidth]{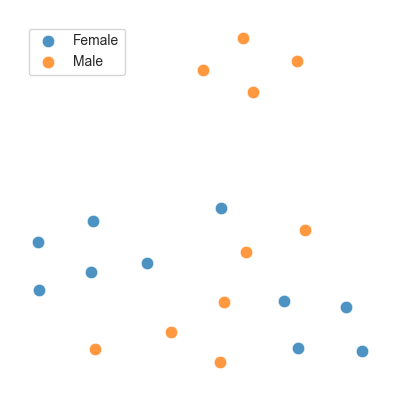}
        \caption{CosyVoice}
        \label{fig:cosyvoice-gender-var}
    \end{subfigure}
    \hfill
    \hfill
    \begin{subfigure}[b]{0.22\textwidth}
        \centering
        \includegraphics[width=\linewidth]{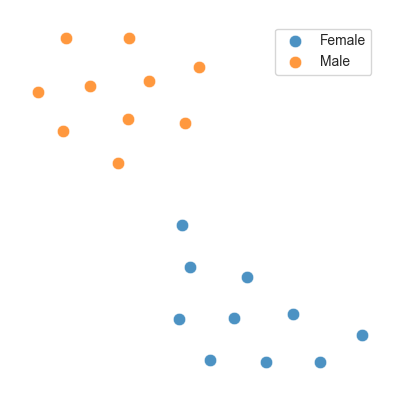}
        \caption{Emotion}
        \label{fig:vibe-gender-var}
    \end{subfigure}
    
    \caption{t-SNE visualizations of the age embedding for speech generated using a fix prompt}
    \label{fig:embed_cosy}
\end{figure}

ParaStyleTTS consistently generates speech with clearly distinguishable gender characteristics across all phrasings. As shown in Figure~\ref{fig:vibe-gender-var}, embeddings of Male and Female speech form two well-separated clusters, indicating that the model maintains stable control regardless of prompt formulation.
CosyVoice, on the other hand, shows weak robustness in this setup. While it performs reliably on Female prompts, it fails to maintain consistency for Male prompts: 5 out of 10 male-prompted samples are perceptually identified as female. This inconsistency is reflected in Figure~\ref{fig:cosyvoice-gender-var}, where multiple Male samples are mis-clustered near the Female region. It highlights CosyVoice's failure to robustly represent style under prompt variation. CosyVoice relies on LLMs to interpret the semantic information from prompts and infer speaking styles. This text-to-style pathway maps semantic meaning to acoustic characteristics through multiple indirect, black-box stages, including semantic interpretation, content-style fusion, and waveform generation. As a result, even small changes in prompt phrasing can cause fluctuations in the LLM's latent representations, leading to unintended variations in the generated speech. Since CosyVoice lacks an explicit control mechanism for aligning acoustic output with the intended style, its control becomes brittle and highly sensitive to the prompt formulation.

In contrast, ParaStyleTTS adopts an acoustic-centric, prompt-to-acoustics learning paradigm. It learns to directly associate style prompts with acoustic features through supervised training, enabling a more grounded and consistent style representation. This results in clear and disentangled mappings between the prompt and the generated output, making the system robust to differences in prompt wording. These findings show that ParaStyleTTS not only achieves higher per-class accuracy across various paralinguistic styles but also maintains robust and consistent style expression under varied prompt formulations. This level of robustness is essential for real-world TTS applications, especially in interactive or open-ended environments, where users may express the same speaking style with a different prompt formulation.

\section{Conclusion}
In conclusion, we propose a novel style-controllable TTS model, ParaStyleTTS, that enables efficient, robust, and expressive control over speaking style. ParaStyleTTS introduces a novel two-level style modeling architecture that captures both local prosodic and global paralinguistic styles and supports flexible control over speaking styles such as emotion, gender, and age. It adopts an acoustic-centric, end-to-end design that can generate high-quality speech directly from input text and style prompts.
Experiments demonstrate that ParaStyleTTS outperforms LLM-based baselines in both style accuracy and computational efficiency. It achieves over 30x faster inference, up to 8x smaller model size, and 2.5x lower memory usage compared to CosyVoice, while maintaining consistent and expressive style control.

\section{Limitations and Future Work}
While ParaStyleTTS demonstrates strong performance in naturalness and paralinguistic style control, it still falls slightly behind CosyVoice in overall intelligibility and subjective naturalness. In future work, we plan to expand the training dataset by incorporating a greater variety of speakers, speaking styles, and languages to help bridge this gap. Additionally, the current model supports only three paralinguistic styles. We aim to extend controllability to a broader range of paralinguistic styles, such as personality, speaking tone, and energy level, to enable a more comprehensive control of speaking style in TTS model.

\begin{table}[t]
\centering
\caption{Distribution of Training Data by Speaking Style}\label{tab:dataset}
\begin{tabular}{llrr}
\toprule
\textbf{Category} & \textbf{Value} & \textbf{Count} & \textbf{Hours} \\
\midrule
\multirow{2}{*}{Gender} 
  & Female     & 51,646 & 69.68 \\
  & Male       & 34,360 & 38.65 \\
\midrule
\multirow{4}{*}{Age} 
  & Child      & 1,463 & 1.90\\
  & Teenager   & 9,107 & 13.77\\
  & Youngadult & 9,012 & 13.25\\
  & Adult      & 66,424& 79.42\\
\midrule
\multirow{4}{*}{Emotion} 
  & Angry      & 7,062 & 5.37\\
  & Sad        & 7,005 & 6.84\\
  & Neutral    & 57,972& 84.89\\
  & Happy      & 6,560 & 5.00 \\
  & Surprise   & 7,407 & 6.24 \\
\midrule
\multirow{2}{*}{Language} 
  & Chinese    & 41,475 & 47.73\\
  & English    & 44,531 & 60.60\\
\bottomrule
\end{tabular}
\end{table}

\begin{table}[ht]
\centering
\caption{Text Prompts Used for Gender-specific Style Control}\label{tab:text_prompt}
\begin{tabular}{|c|p{7cm}|}
\hline
\textbf{Gender} & \textbf{Prompt} \\
\hline
Female & A female speaker is talking. \\
Female & You are listening to a woman speak. \\
Female & This voice belongs to a female speaker. \\
Female & A young woman is speaking in a calm tone. \\
Female & A woman is narrating this sentence. \\
Female & A girl is speaking softly. \\
Female & You're hearing the voice of a lady. \\
Female & A lady is giving this speech. \\
Female & This is the voice of a female child. \\
Female & A girl is talking in Chinese. \\
\hline
Male & A male speaker is talking. \\
Male & You are hearing a man's voice. \\
Male & This voice belongs to a male speaker. \\
Male & A man is speaking in a confident tone. \\
Male & A male narrator is delivering the sentence. \\
Male & A man is speaking cheerfully. \\
Male & You're hearing the voice of a gentleman. \\
Male & A gentleman is giving this speech. \\
Male & This is the voice of a male. \\
Male & A gentleman is talking in Chinese. \\
\hline
\end{tabular}
\label{tab:gender_prompts_vertical}
\end{table}
\section{GenAI Usage Disclosure}
During the development and writing of this paper, generative AI (GenAI) tools are used in limited, non-substantive ways. Specifically, we use GenAI tools to help refine grammar, improve clarity, and restructure paragraphs in the manuscript. All technical content, research insights, and model designs are solely authored by the authors without any GenAI-generated ideas. No GenAI tools are used to generate or modify source code, experiment design, or model training. All implementation and data handling are conducted manually using standard Python-based, PyTorch frameworks. All datasets used are publicly available human speech datasets. Data preprocessing and analysis are performed without the aid of GenAI tools.

\bibliographystyle{ACM-Reference-Format}
\balance
\bibliography{custom}
\end{document}